\documentclass{Interspeech2024}
\usepackage{multirow}

\interspeechcameraready

\title{Towards Audio Codec-based Speech Separation}

\name[affiliation={1,3}]{Jia Qi}{Yip}
\name[affiliation={3}]{Shengkui}{Zhao}
\name[affiliation={2,3}]{Dianwen}{Ng}
\name[affiliation={2}]{Eng Siong}{Chng}
\name[affiliation={3}]{Bin}{Ma}

\address{
    $^1$Alibaba-NTU Singapore JRI, Interdisciplinary Graduate Programme, NTU, Singapore \\
    $^2$Nanyang Technological University, Singapore\\
    $^3$Alibaba Group
  }
\email{jiaqi006@e.ntu.edu.sg}

\keywords{speech separation, audio codec, resource efficient, neural audio compression}

\begin{document}

\maketitle{}

\begin{abstract}
    Recent improvements in neural audio codec (NAC) models have generated interest in adopting pre-trained codecs for a variety of speech processing applications to take advantage of the efficiencies gained from high compression, but these have yet been applied to the speech separation (SS) task. SS can benefit from high compression because the compute required for traditional SS models makes them impractical for many edge computing use cases. However, SS is a waveform-masking task where compression tends to introduce distortions that severely impact performance. Here we propose a novel task of Audio Codec-based SS, where SS is performed within the embedding space of a NAC, and propose a new model, Codecformer, to address this task. At inference, Codecformer achieves a 52x reduction in MAC while producing separation performance comparable to a cloud deployment of Sepformer. This method charts a new direction for performing efficient SS in practical scenarios.
\end{abstract}

\section{Introduction}
Speech Separation is the task of obtaining the audio of separated speakers from a mixture of speakers. It is also often known as the cocktail party problem~\cite{cherry1953some}, which was the observation that humans are able to efficiently pay attention to audio sources in a noisy environment, b  ut this task is difficult for computers. Speech Separation is trained on waveform comparison loss against ground truth clean speech, typically using the Scale-invariant Signal-to-Distortion Ratio (SI-SDR)~\cite{le2019sdr} loss function. Due to this stringent loss, it has been found that performance typically suffers as compression of the input audio increases~\cite{Subakan2022ExploringSM}. As a result, recent models have been growing in size and computational cost, especially with models that combine both time and frequency domain features~\cite{zhao2023mossformer2}. 

Neural Audio Compression (NAC) seeks to find compressed representation of audio that can be reconstructed with good fidelity compared to the original audio~\cite{zeghidour2021soundstream}. These models are often trained using generative adversarial networks, where a discriminator model seeks to determine if any given audio sequence is generated or not. In addition, a number of waveform level comparison losses like SI-SDR are also used to measure the model output against the original audio.

\begin{figure}[t!]
  \centering
  \includegraphics[width=\linewidth]{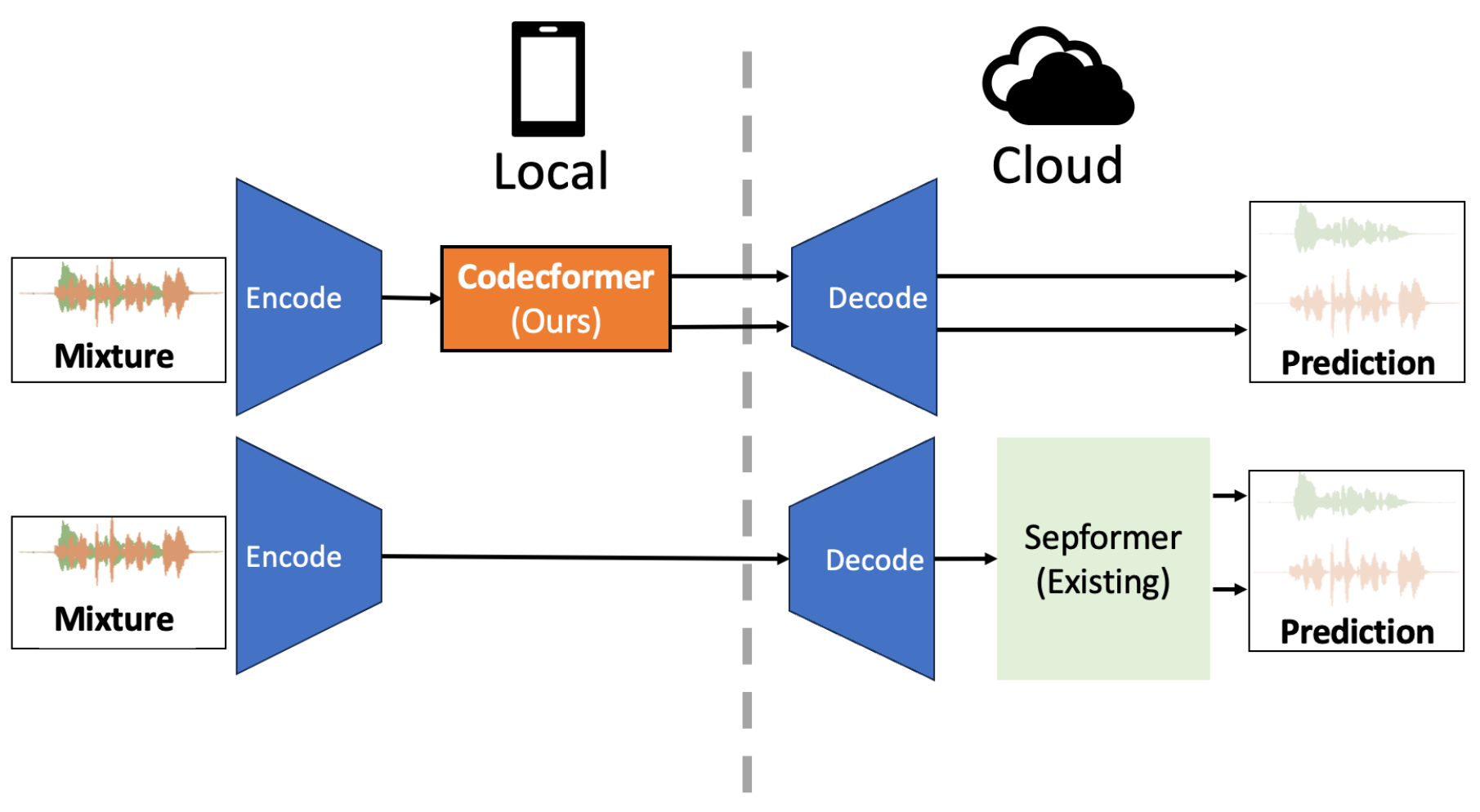}
  \caption{(Top) Overview of Codecformer performing speech separation within the embedding space of an audio codec. The decoding can happen either locally or in the cloud. (Bottom) The existing approach of performing speech separation in the cloud after transmitting a mixture through a codec}
  \label{fig:codecformer-overview}
\end{figure}

Due to the high compression achieved by the discrete audio representations generated by neural audio codecs, they have been used in a variety of applications such as Speaker Verification, Diarization and Speech Recognition~\cite{puvvada2023discrete}. They have also been used as an audio-encoder for multi-modal large language models~\cite{wang2023speechx}~\cite{Wang2023LauraGPTLA} as an alternative to Whisper~\cite{radford2023whisper}-based continuous embedding approaches~\cite{chu2023qwen} or HuBERT~\cite{hubert}-based approaches~\cite{zhang-etal-2023-speechgpt}. These experiments show that despite a high level of compression, a significant portion of information and a wide variety of features contained within the audio can be captured by the neural codecs. This makes NACs an attractive approach for performing speech separation in edge computing situations.

Performing speech separation using codecs has not been attempted before this because of a number of difficulties. Firstly, since audio of speaker mixtures is typically not part of the training set, the codecs may not be able to represent mixtures of voices in its compressed embedding space. Secondly, speech separation performance is typically measured through waveform matching which would be very sensitive to the distortions created by the compression~\cite{wang2018supervised}. Similar attempts using text-to-speech instead of codecs have relied non-waveform matching methods and perceptual quality metrics~\cite{shi2021discretization}. 

Nevertheless, speech separation can benefit greatly from operating in the compressed space of neural audio codecs. State-of-the-art speech separation models~\cite{zhao2023mossformer2}~\cite{sepformer}~\cite{SPGM} are difficult to deploy in edge computing use cases because they are large and have high computational requirements. Typically, audio will have to be sent to the cloud for processing, which may have to be transmitted through a codec anyway, as shown in Figure~\ref{fig:codecformer-overview}. Additionally, using audio codecs pre-trained on massive amounts of data has the potential to benefit the speech separation task during fine-tuning.

\subsection{Our Approach}
As shown in Figure~\ref{fig:codecformer-overview}, instead of the traditional approach of using a full audio sequence to generate the speech separation output, our separation model accepts only the embeddings of an audio codec. This embedding dimension is significantly smaller than the typical encoders used in speech separation models. One reason is that neural codecs perform significant time compression, reducing sequence lengths significantly. This allows our proposed model, Codecformer\footnote{https://github.com/Yip-Jia-Qi/codecformer}, to accept the full sequence into its transformer layers without the need for any chunking. It also significantly reduces the memory and computational requirements of training the model. 

In this work we discuss the appropriate ground truth for this training setting, proposing the Codec SI-SDR loss metric as a new target for such systems. Our proposed model, Codecformer, utilizes 52x fewer MACs at inference compared to Sepformer while producing similar separation quality. To our knowledge, this is the first attempt at performing separation on the intermediate features of an audio codec.

\section{Methodology}
\subsection{Codecformer}
\begin{figure}[t!]
  \centering
  \includegraphics[width=\linewidth]{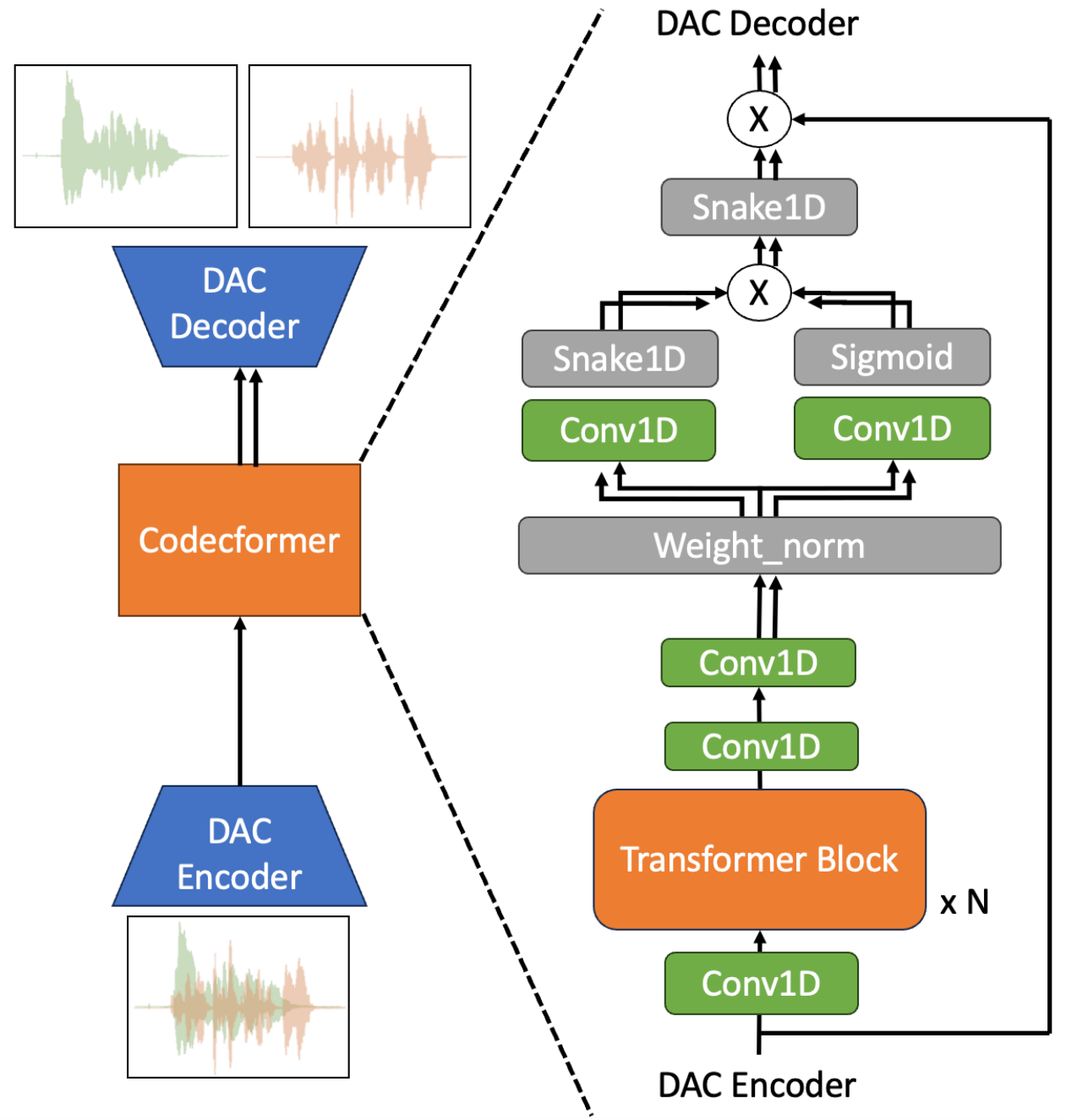}
  \caption{(Left) Overview of Codecformer within the context of the DAC Encoder and Decoder. In this work we use DAC, but this can be replaced with any other neural audio codec. (Right) The detailed implementation of Codecformer. The transformer block is identical to that of  Sepformer~\cite{sepformer}}
  \label{fig:separator-architecture}
  \vspace{-10pt}
\end{figure}

The detailed architecture of our separation model, Codecformer, is shown in Figure~\ref{fig:separator-architecture}. The model is based on the popular Sepformer~\cite{sepformer} architecture, but modified to work within the embedding space of the Descript Audio Codec (DAC)~\cite{DACkumar2024high}, discussed in Section~\ref{sec:dac}, which is the NAC used for this work.

Firstly, to manage the long sequence lengths, models such as Sepformer~\cite{sepformer} also have to resort to chunking methods and a dual-path architecture~\cite{Luo2019DualPathRE} as the full audio sequence cannot be fed into a vanilla transformer without exceeding memory limitations. Due to the high temporal compression provided by DAC, this limitation is lifted. This allows us to simplify our model architecture by removing the need for Intra and Inter blocks in the typical dual-path separation approach, replacing them with a simple stack of transformer layers that could be configured to be compatible with a real-time continuous communication system.

Secondly, the model utilizes the Snake activation function~\cite{snake} instead of the tanh and ReLU functions used in speech separation models. The snake activation is written as $x+\sin^{2}(x)$ and is designed to be better at modeling periodic functions by inducing a periodic bias. In our early ablation studies we tried a number of activation functions, including tanh and ReLU but the performance was very poor. We suspect that this could be because DAC makes use of the Snake activation and it was necessary for the separation model's final activation layer to match that of the codec so that the magnitude of the values returned are similar.

\subsection{Descript Audio Codec}
\label{sec:dac}
In this work we make use of the Descript Audio Codec~\cite{DACkumar2024high} (DAC), which is a recently released open source model which improves on EnCodec~\cite{encodec}. We adopt the DAC model because of its strong performance relative to other NACs as well as its training dataset consisting of a diverse range of data, including both speech, audio and music. DAC consists of three main components, an Encoder, a Decoder and a Residual Vector Quantization (RVQ) block that is used for transmission.

The model implementation and weights are used as provided by the authors\footnote{https://github.com/descriptinc/descript-audio-codec}. However, a wrapper is used to integrate DAC with our training framework Speechbrain~\cite{speechbrain}. 

\subsection{Loss Functions}
In order to train Codecformer, the desired target loss function needs to be adjusted. In addition to the standard SI-SDR, we propose an additional loss function, Codec SI-SDR, to guide the training of Codecformer, where the difference is that the ground-truth used includes the distortions created by the codec.
\subsubsection{SI-SDR}
The SI-SDR loss function as described by~\cite{le2019sdr} can be written as follows:
\begin{equation}
\begin{aligned}
    \text{SI-SDR} &= 10\text{log}_{10}\frac{||\alpha s||^{2}}{||\alpha s - \hat{s}||^{2}} \\
                  & \text{where } \alpha = \frac{\hat{s}^{T}s}{||s||^2}
   \end{aligned}
\end{equation}

SI-SDR is a commonly used metric for evaluating the performance of speech separation models. It measures the improvement in signal-to-noise ratio (SNR) achieved by the model, normalized to account for variations in the overall signal level. A higher SI-SDR value indicates better performance, with 0 dB signifying no improvement and negative values indicating that the estimated signal is worse than the original mixture.

\subsubsection{Codec SI-SDR}
Our proposed Codec SI-SDR formula can be written as follows:
\begin{equation}
\begin{aligned}
    t &= \text{Codec}(s) \\
    \text{cSI-SDR} &= 10\text{log}_{10}\frac{||\alpha t||^{2}}{||\alpha t - \hat{s}||^{2}} \\
                  & \text{where } \alpha = \frac{\hat{s}^{T}t}{||t||^2}
   \end{aligned}
\end{equation}
The standard SI-SDR loss function, as described in the previous section, compares the original ground truth $s$ with the estimated signal $\hat{s}$ after scaling them by a factor $\alpha$. However, this approach may not provide a fair comparison because in an actual deployment context the cloud would also be receiving a transmission of the original ground truth, after some distortion. We propose a modified version called ``Codec SI-SDR" (cSI-SDR) that compares the estimated signal $\hat{s}$ against the transmitted signal $t$ obtained by encoding a clean signal into codes locally and decoding the codes on the server. This modification allows us to account for the additional reconstruction errors introduced by the codec. 


\section{Experiments}

\subsection{Dataset}
In all our experiments we make use of the popular benchmark dataset WSJ0-2mix~\cite{wsj0mix}. The dataset consists of mixtures combined using clean speech from the WSJ0 dataset. Both mixtures and the ground truth consist of speech recorded in acoustically treated environments at an 8kHz sampling rate. The WSJ0-2Mix dataset consists of mixtures drawn from the WSJ0 corpus and consists of 20,000 mixtures (30 hours) from the WSJ0 si\_tr\_s set, a validation set of 5000 mixtures (10 hours) from the WSJ0 si\_dt\_05 set and a test set of 3000 mixtures (5 hours) from the si\_et\_05 set.

\subsection{Hyperparameters}
The DAC model uses embedding size 1024, and where applicable the full-bitrate of the DAC model was used. Codecformer was set up with an embedding size of 256 and 16 transformer blocks. The number of transformer blocks were chosen based on the design of Sepformer, which has 8 intra blocks and 8 inter blocks for a total of 16 blocks. Training was performed with an initial learning rate of 1.5$e_{-4}$ and the LR scheduler was set to halve the learning rate with a patience of 2 after epoch 5. All models were trained to 200 epochs. 

\subsection{Baselines}
We report in Table~\ref{tab:ablation} three baselines to benchmark the performance of a popular speech separation model, Sepformer~\cite{sepformer}, in various usage scenarios. The model is pre-trained and implemented by the Speechbrain~\cite{speechbrain} framework.

\noindent \textbf{Sepformer (Oracle)} is the baseline where no codec is used. The input to Sepformer is the original mixture and the output of Sepformer is used directly in the performance comparison. This allows us to measure the distortion with respect to the degraded transmissions. 

\noindent \textbf{Sepformer (Local)} where the model is given a clean mixture, but the output is transmitted through DAC, resulting in distortions. This mimics a scenario where the separation is performed on-device and its output is sent to the cloud. 

\noindent \textbf{Sepformer (Cloud)} where the model is given a distorted mixture that has been transmitted through DAC. This implements the scenario shown at the bottom of Figure~\ref{fig:codecformer-overview}.

\section{Results}
In Table~\ref{tab:simplecomparison} we compare the results of Codecformer against the Sepformer (Cloud) baseline. The Sepformer (Cloud) baseline represents the performance of the model if it was necessary to transmit the mixture to the cloud for processing. Meanwhile, Codecformer performs the separation locally. Since we are comparing transmitted audio, we append ``c'' to the performance metrics SDRi and SI-SDRi to denote that although we use the same computation, the basis of comparison is the ground-truth that has been distorted by the codec. In this comparison, we see that Codecformer outperforms Sepformer (Local) by 0.3dB on both cSI-SDRi and cSDRi, and is competitive on perceptual quality based on PESQ. Nevertheless, we believe that these objective metrics may understate the true performance of Codecformer, thus we have also provided audio samples for the reader to make a subjective evaluation\footnote{https://github.com/Yip-Jia-Qi/codecformer/tree/main/samples}.

\renewcommand{\arraystretch}{1.2}
\begin{table}[h!]
  \caption{Comparison of performance between Sepformer (Cloud) and Codecformer trained using cSI-SDRi}
  \label{tab:simplecomparison}
  \centering
    \begin{tabular}{ |c|c|c|c|  }
    \hline
    Model & cSI-SDRi & cSDRi & PESQ \\
    \hline
    Sepformer (Cloud) & 9.6 & 10.1 & \textbf{2.75}\\
    Codecformer (Ours) & \textbf{9.9} & \textbf{10.4} & 2.58 \\
    \hline
    \end{tabular}
\end{table}

\subsection{Improvements to Training Speed and Computation}
The main advantage of performing speech separation within the embedding space of a NAC is the reduced computation required and improved training speed, which we report in Table~\ref{tab:compandspeed}. Comparing against Sepformer training, Codecformer achieves a 2.7x improvement in training speech in terms of hours per epoch, which can translate to significant cost savings. These numbers are obtained on a machine with a V100 with 16GB of RAM. For practical deployment, we compute Multiple and Accumulate operations (MACs) using PyTorch-OpCounter\footnote{https://github.com/Lyken17/pytorch-OpCounter} to measure the inference requirements for the models in a hardware agnostic manner. In this case the test input utterance is 2 seconds of 8kHz audio and the MACs from the audio compression and decompression is not included since it would apply to both separators. From the results in Table~\ref{tab:compandspeed} we can see that a Codecformer requires significantly lower MACs, constituting a 52x reduction to process the same length of audio. This significant improvement is largely due to the time compression afforded by DAC, which reduces the 8kHz audio to 50Hz. As the shorter sequence lengths reduces computational requirements.

\renewcommand{\arraystretch}{1.2}
\begin{table}[h!]
  \caption{Computation and traning speech of Codecformer compared against Sepformer}
  \label{tab:compandspeed}
  \centering
    \begin{tabular}{ |c|c|c|  }
    \hline
    Model & GMACs & Training Time (h/epoch) \\
    \hline
    Sepformer~\cite{sepformer} & 77.3 & 2.7\\
    Codecformer (Ours) & \textbf{1.5} & \textbf{1.0}\\
    \hline
    \end{tabular}
    \vspace{-10pt}
\end{table}

\renewcommand{\arraystretch}{1.2}
\begin{table*}[t!]
\vspace{-5pt}
  \caption{Ablation study of different training targets for Codecformer, compared against different Sepformer baselines.}
  \vspace{-5pt}
  \label{tab:ablation}
  \centering
    \begin{tabular}{ |p{3cm}|c||c|c|c|c||c|c|c|c|c|  }
     \hline
    \multicolumn{2}{|c||}{} & 
    \multicolumn{4}{c||}{\textbf{Ground Truth}} & 
    \multicolumn{4}{c|}{\textbf{Transmission}} & \multicolumn{1}{c|}{\multirow{2}{*}{\textbf{PESQ}}} \\ \cline{1-10} 

    \multicolumn{2}{|c||}{\textbf{Baselines}} & \multicolumn{1}{c|}{\textbf{SI-SDR}} & \multicolumn{1}{c|}{\textbf{SI-SDRi}} & \multicolumn{1}{c|}{\textbf{SDR}} & \multicolumn{1}{c||}{\textbf{SDRi}} & \multicolumn{1}{c|}{\textbf{cSI-SDR}} & \multicolumn{1}{c|}{\textbf{cSI-SDRi}} & \multicolumn{1}{c|}{\textbf{cSDR}} & \multicolumn{1}{c|}{\textbf{cSDRi}} & \multicolumn{1}{c|}{} \\ \hline
    \multicolumn{2}{|c||}{Sepformer (Oracle)} & 22.4 & 22.4 & 22.8 & 22.6 & -20.6 & -19.5 & 12.3 & 13.0 & 4.00\\
    \multicolumn{2}{|c||}{Sepformer (Local)} & -20.6	& -20.6	& 0.2	& 0.1 & 11.6 & 12.7	& 12.3 & 13.0 & 3.84\\
    \multicolumn{2}{|c||} {Sepformer (Cloud)} & -20.7 & -20.7 & 0.0 & -0.2 & 8.6 & 9.6 & 9.3 & 10.1 & 2.75\\
    \hline

    \multicolumn{1}{|c|}{\textbf{Training Target (Loss)}} & \multicolumn{1}{c||}{\textbf{RVQ}} & \multicolumn{9}{c|}{} \\ \hline
    
    \multirow{2}{*}{Transmission (cSI-SDR)} & \checkmark & -20.6 & -20.6 & -0.6 & -0.8 & 5.9 & 7.0 & 6.9 & 7.6 & 2.07 \\
     & $\times$ & -20.5 & -20.5 & 0.1 & -0.1 & \textbf{8.8} & \textbf{9.9} & \textbf{9.7} & \textbf{10.4} & \textbf{2.58}\\
    \hline
     \multirow{2}{*}{Ground Truth (SI-SDR)}& \checkmark & 5.0 & 5.0 & 6.4 & 6.2 & -20.5 & -19.4 & 6.6 & 7.3 & 2.09\\
     & $\times$ & \textbf{8.7}	& \textbf{8.7} & \textbf{9.8} & \textbf{9.7} & -20.6 & -19.6	& 9.4 & 10.1 & \textbf{2.52}\\
    \hline
    \end{tabular}
    
\end{table*}

\subsection{Ablation study}
In Table~\ref{tab:ablation} we report the results of ablation studies over the different methods of training Codecformer. For the column labeled ``Ground Truth'', we use the original clean speech as the basis for comparison. Next, for the column labeled ``Transmission'', we use degraded speech produced by transmitting the original clean speech through DAC as the basis for comparison. In both cases, we report all 4 commonly used performance metrics for speech separation (with ``c'' denoting the use of transmission as the basis for comparison), in addition to PESQ to measure the perceptual quality of the output. In this study, PESQ is measured against the original ground truth since it would not be principled to use the distorted audio as the comparison.
\vspace{-5pt}
\subsubsection{Effects Transmission Distortions}
Typically, transmitting audio through a codec results in distortions. This can result in poor performance on the objective metrics but the audio can still perform well on the perceptual metric. In the Sepformer (Oracle) baseline, we see that while the model's output achieves good performance against the original ground truth, but when the basis of comparison is the transmission, the performance drops significantly although perceptual quality remains high. Conversely, in the Sepformer (Local) baseline, we see that PESQ suffers only a small drop from 4.00 to 3.84 even through the comparison against the ground truth results in poor performance. Additionally, we note that the scale-invariant metrics suffer significantly whenever there is a comparison between a degraded and original waveform. This is likely a scaling issue from the distortions introduced by DAC.

\vspace{-5pt}
\subsubsection{Effects of Training Target}
Comparing the results of Codecformer trained on the transmission target against those trained on ground truth target, we see that training on the ground truth benefits the model more broadly, delivering performance under both ground truth and transmission comparisons. Training on a transmission target only benefits the model on the transmission comparison and causes poor performance when compared against the ground truth. We believe that this is because using the ground truth signal provides a cleaner loss gradient to the model, whereas when training using the transmission target the model may occasionally be penalized for the inherent distortions in the transmission.

\subsubsection{Effects of RVQ}
A common feature of neural audio compression models is the Residual Vector Quantization (RVQ)~\cite{Gray1984VectorQ}\cite{Vasuki2006ARO} layer which compresses model embeddings by learning a quantization codebook to represent the embeddings with only a small number of codes. The first layer of the RVQ module aims to calculate an estimate of the original embedding and each subsequent layer estimates the residual between the sum of the previous layers and the original embedding. Each subsequent code is thus of decreasing importance to reconstruction, allowing the transmission bitrate to be easily scalable. This design, although efficient, means it is inevitable that the RVQ module results in some degradation of the signal~\cite{puvvada2023discrete}.  

Thus, in addition to experimenting with different targets and losses for training, for each target we perform an ablation study on the use of RVQ in the model. Experiments where RVQ are used show the performance of the separation if the model was run in the cloud. On the other hand, our experiments without the use of RVQ, which show the potential maximum separation performance in the limit of an RVQ block with an infinite number of layers.

We observe the negative effect of RVQ in all of our experimental results, where the experiments with the use of RVQ results in a ~2-3dB reduction in performance across all the experiments. On the perceptual quality metric, RVQ causes degradation as well, reducing PESQ by ~0.5. Overall these results shows that there are important limitations in the use of RVQ for DAC that results in transmission loss, and highlight a potential source of future gains in performance as RVQ methods improve.

\vspace{-5pt}
\section{Conclusion}
In this work we have explored a promising approach towards speech separation within a NAC and proposed a model, Codecformer to tackle this task. Codecformer achieves a reduction in the MAC required during inference by 52x compared to the baseline Sepformer architecture and trains 2.7x faster while producing competitive performance on both objective and perceptual metrics. Thus, our results have shown that the proposed task of performing speech separation within the embedding space of a NAC is indeed possible, allowing us to gain inference and memory advantages over traditional speech separation approaches. 

One limitation of this work is that DAC was not explicitly trained with speech mixtures in the training dataset, which could result in its internal representation lacking the capability to represent important features in fully-overlapping speech. Future work along this domain could include pre-training a NAC on mixture data for improved performance, as well as exploring the use of Codecformer as a first-cut model for downstream applications like diarization, ASR or speaker extraction. Codecformer can also suffice for applications that need to prioritize inference speed. Additionally, there is also an opportunity here to perform speech separation as a token-level auto-regressive task, allowing us to use cross-entropy as a loss function.

\vspace{-5pt}
\section{Acknowledgements}
This research is supported by the RIE2025 Industry Alignment Fund – Industry Collaboration Projects (IAF-ICP) (Award I2301E0026), administered by A*STAR, as well as supported by Alibaba Group and NTU Singapore. We would like to acknowledge Alibaba-NTU Joint Research Institute, Interdisciplinary Graduate Programme, Nanyang Technological University, Singapore
\bibliographystyle{IEEEtran}
\bibliography{mybib}

\end{document}